# Fractal Self-Assembled Nanostructures on Monocrystalline Silicon Surface

N.T.Bagraev[1a], A.D.Bouravleuv[1a], W.Gehlhoff[2b], L.E.Klyachkin[1a], A.M.Malyarenko[1a], V.V.Romanov[3a] and S.A.Rykov[3a]

[1]A.F.Ioffe Physico-Technical Institute, St.Petersburg, Russia

[2]Institut fuer Festkoerperphysik, Technische Universitaet Berlin, Berlin, Germany

[3]St.Petersburg State Technical University, St.Petersburg, Russia

[a]impurity.dipole@mail.ioffe.ru, [b]gehlhoff@sol.physik.tu-berlin.de

**Abstract.** We present ultra-shallow diffusion profiles performed by short-time diffusion of boron from the gas phase using controlled surface injection of self-interstitials and vacancies into the n-type Si(100) wafers. The diffusion profiles of this art are found to consist of both longitudinal and lateral silicon quantum wells of the p-type that are self-assembled between the alloys of microdefects, which are produced by previous oxidation. These alloys appear to be passivated during short-time diffusion of boron thereby forming neutral $\delta$ - barriers. The fractal type self-assembly of microdefects is found to be created by varying the thickness of the oxide overlayer, which causes the system of microcavities embedded in the quantum well plane.

## Introduction

Dopant diffusion in semiconductors is amenable to be controlled by means of adjusting the fluxes of self-interstitials and vacancies emerging from the silicon monocrystalline surface [1]. In order to increase the rates of diffusion for shallow impurities, which penetrate into the silicon lattice by the kick-out and vacancy diffusion mechanisms, it is necessary to use thin oxide overlayers in combination with high diffusion temperatures [1, 2] as well as the package of thick oxide overlayers and low diffusion temperatures [2] that result respectively in the generation of self-interstitials and vacancies by the oxidized surface. Hence, the retardation of the diffusion process achieved under parity between the kick-out and vacancy mechanisms for the formation of ultra-shallow boron and phosphorus diffusion profiles into silicon wafers has to be clarified by varying the oxide overlayer thickness and the diffusion temperature [2, 3]. However, the impurity diffusion in the process of oxidation is accompanied by the formation of point centres and microdefects, which prevent the preparation of ultra-shallow *p-n* junctions that possess low leakage currents. The goal of the present work is to use short-time diffusion of boron as the passivation of the self-assembled alloys of microdefects produced previously by the oxidation of the Si(100) wafers for the preparation of ultra-narrow silicon quantum wells and two-dimensional microcavities.

## Self-assembled alloys of microdefects in the Si(100) wafers

The preparation of oxide overlayers on silicon monocrystalline surfaces is known to be favourable to the generation of the excess fluxes of self-interstitials and vacancies that exhibit the predominant crystallographic orientation along a <111> and <100> axis, respectively (Fig. 1a) [2, 4]. In the initial stage of the oxidation, thin oxide overlayer produces excess self-interstitials that are able to create small microdefects, whereas oppositely directed fluxes of vacancies give rise to their annihilation (Figs. 1a and 1b). Since the points of outgoing self-interstitials and incoming vacancies appear to be defined by the positive and negative charge states of the reconstructed silicon dangling bond [4], the dimensions of small microdefects of the self-interstitials type near the Si(100) surface have to be restricted to 2 nm. Therefore, the distribution of the microdefects created at the initial stage of the oxidation seems to represent the fractal of the Sierpinski Gasket type with the built-in longitudinal silicon quantum well (LSQW) (Figs. 1b and 2a). Then, the fractal distribution has to be reproduced by increasing the time of the oxidation process, with the $P_b$ centres as the germs for the next generation of the microdefects (Fig. 1c and 2b) [5]. The formation of thick oxide overlayer under prolonged oxidation results in however the predominant generation of vacancies by the oxidized surface, and thus, in increased decay of these microdefects, which is accompanied by the self-assembly of the lateral silicon quantum wells (LaSQW) (Figs. 1d and 2c). Although both LSQW and LaSQW embedded in the fractal system of self-assembled microdefects are of interest to be used as a basis for optically and electrically active microcavities in optoelectronics and nanoelectronics, the presence of dangling bonds at the interfaces prevents such an application. Therefore, subsequent short-time diffusion of boron would be appropriate for the passivation of dangling bonds and other defects created during previous oxidation of the Si(100) wafers thereby assisting the transformation of the alloys of microdefects in neutral δ - barriers confining self-assembled silicon quantum wells (Figs. 2a', 2b', 2c', 3 and 4).

## Methods

The 0.35 mm thick n- type Si(100) wafers with resistivity 20 Ohm·cm were previously oxidized at 1150°C in dry oxygen containing $CCl_4$ vapors. The thickness of the oxide overlayer is dependent on the duration of the oxidation process that was varied from 20 min up to 24 hours. The short-time diffusion of boron was done during five minutes into windows which were cut in the oxide after preparing a mask and performing the subsequent photolithography. Additional replenishment with dry oxygen into the gas phase during the diffusion process provides the generation of excess fluxes of intrinsic point defects from the working side. The variable parameters of the diffusion experiment were the oxide overlayer thickness, the diffusion temperature and the Cl levels in the gas phase during the diffusion process. Diffusion profiles were studied using the four-point probe, SIMS, Cyclotron Resonance (CR), Infrared Fourier spectroscopy and Scanning Tunneling Microscopy (STM) techniques.

## Results

Figs. 2 a, b and c show the important role of the diffusion temperature and the thickness of the oxide overlayer in determining the diffusion profile depth for the kick-out and vacancy diffusion mechanisms. Since the diffusion process proceeds at 800°C and 1100°C respectively via the vacancy and the kick-out mechanisms, the smallest penetration depths were observed under parity that is revealed by the diffusion temperature of 900°C (Fig. 2b). Besides, the retardation of the diffusion process is found to result in the shallowest diffusion profile in the presence of thin oxide overlayer when the fluxes of boron from the gas phase are able to passivate only the alloys of microdefects (Fig. 2a and 2a') without the penetration into bulk through the vacancy-related spikes. The diffusion profile depth is seen to reach the maximum at the diffusion temperature of 800°C and 1100°C (Figs. 5a and 5c), which is due to the enhancement of the diffusion of boron respectively through the vacancy-related spikes and the exchange with self-interstitials coming from microdefects (Fig. 2a,b and 2a',b').

The different depth of the diffusion profiles studied seems to result from the previous formation of LSQW (Figs. 2a' and 5b) and LaSQW (Figs. 2b',c' and 5a) that are identified in such structures by the cyclotron resonance (CR) angular dependencies [6]. These CR measurements were performed at 3.8 K with an EPR spectrometer at X-band (9.1-9.5 GHz). The rotation of the magnetic field in a plane normal to the diffusion profile plane has revealed the anisotropy of both the electron and hole effective masses in silicon bulk and Landau levels scheme in SQW. This CR quenching and the line shifts for which a characteristic 180° symmetry was observed can be explained with the effect of the electrical field created by the confining potential inside $p^+$-diffusion profile and its different arrangement in LSQW and LaSQW formed naturally between neutral δ barriers created by the boron passivation of self-assembled alloys of microdefects. The results obtained exhibit the LSQW into the diffusion profiles prepared at 900°C (Fig. 5 b), while the lateral SQW cross the 800°C structures (Fig. 5a).

The self-assembled alloys of microdefects that cause the formation of SQW are revealed by the STM method as the deformed potential fluctuations (DPF) after etching the oxide overlayer and after subsequent short-time diffusion of boron. The DPF effect induced by the microdefects of the self-interstitials type that are displayed as light poles in Figs. 6,7 and 8 is find to be brought about by the previous oxidation and to be enhanced by subsequent boron diffusion [3]. The STM images demonstrate that the ratio between the dimensions of microdefects produced during the different stages of the oxidation process is supported to be equal to 3.3 thereby defining the self-assembly of microdefects as the self-organization of the fractal type. The analysis of the STM image of the ultra-shallow boron profile prepared under parity conditions between diffusion mechanisms enables to hazard a conjecture that the dimension of the smallest microdefect observed is consistent with the parameters expected from the tetrahedral model of the $Si_{60}$ cluster (Fig. 8) [7]. Finally, the self-assembled alloys of microdefects embedded into the SQW system appear to be a basis of the silicon microcavities (Fig. 9a) that are revealed by the measurements of the reflectance spectra, which exhibit the band structure of a photon crystal (Fig. 9b).

## Summary

The short-time diffusion of boron has been used, for the first time, for the passivation of the self-assembled alloys of microdefects produced previously by the oxidation of the Si(100) wafers. The ultra-shallow diffusion profiles of boron in the Si(100) wafers have been prepared, which consist of the passivated alloys of microdefects that represent the neutral δ - barriers confining silicon quantum wells. The fractal type self-assembly of microdefects has been found to promote the realization of silicon microcavities embedded in the quantum well plane.

## Acknowledgements

This work has been supported by the programme ISTC (Grant #2136) and Personal grant from the President of the Russian Federation supporting young PhD scientists (MK-4092.2004.2).

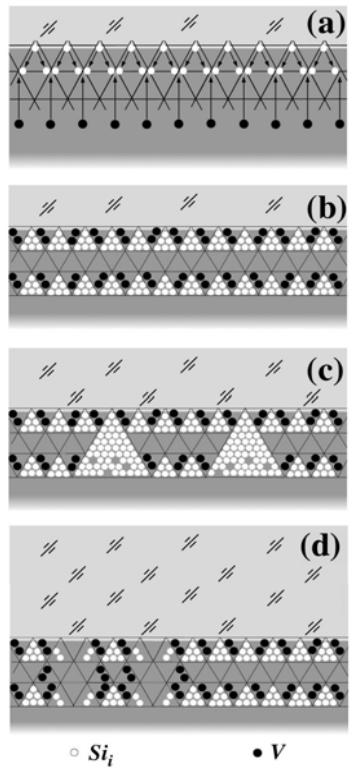

**Fig. 1**. A scheme of self-assembled silicon quantum wells obtained by varying the thickness of the oxide overlayer prepared on the Si(100) wafer. The white and black balls label the self-interstitials and vacancies forming the excess fluxes oriented crystallographically along a <111> and <100> axis that are transformed to small microdefects (a, b). The longitudinal silicon quantum wells between the alloys of microdefects are produced by performing thin oxide overlayer (b), whereas growing thick oxide overlayer results in the formation of additional lateral silicon quantum wells (d). Besides, medium and thick oxide overlayers give rise to the self-assembled microdefects of the fractal type (c).

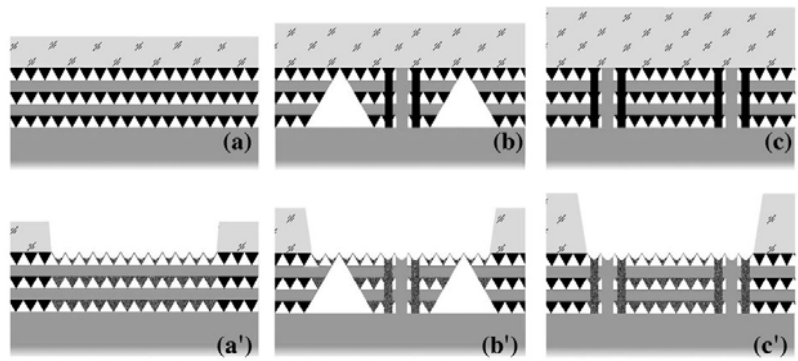

**Fig. 2**. A scheme of ultra-shallow diffusion profiles that consist of the longitudinal (a, a') and lateral (b, b', c, c') quantum wells prepared in the process of the previous oxidation (a, b, c) and subsequent short-time diffusion of boron (a', b', c') on the Si(100) wafer. The atoms of boron replace the positions of vacancies thereby passivating the alloys of microdefects and forming the neutral δ barriers.

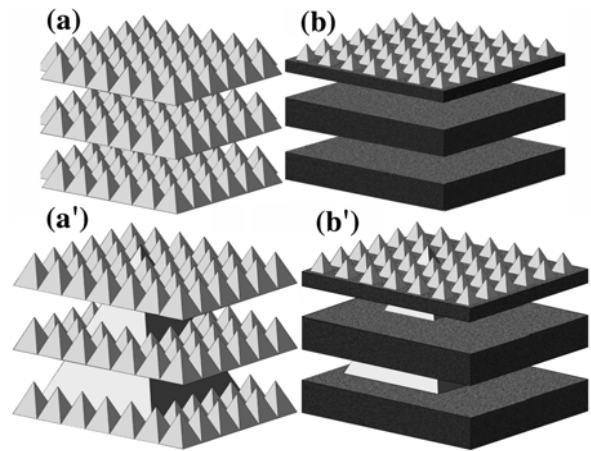

**Fig. 3.** A scheme of longitudinal silicon quantum wells between the alloys of microdefects (a, a')  that are transformed in the neutral δ barriers (b, b') by short-time diffusion of boron. The microdefect of the fractal type that is introduced into the system of the quantum wells is also shown (a', b').

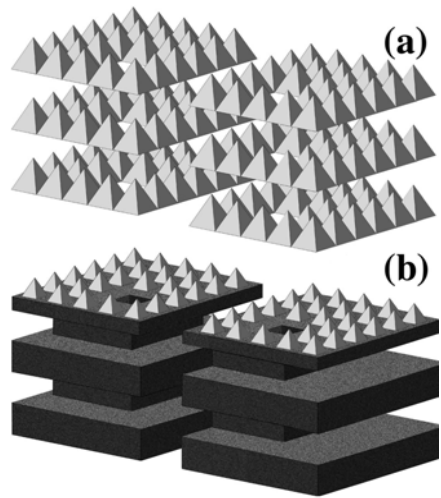

**Fig. 4.** A scheme of lateral silicon quantum well that crosses the system of longitudinal silicon quantum wells between the alloys of microdefects (a), which are transformed in the neutral δ barriers (b) by short-time diffusion of boron. The swirles doped with boron are also shown (b).

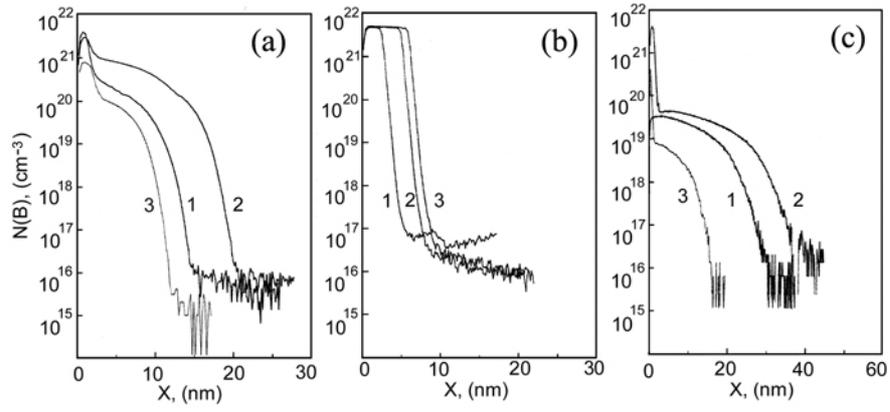

**Fig. 5**. SIMS data for boron diffusion profiles obtained at diffusion temperatures of 800°C (a), 900°C (b) and 1100°C (c) into the Si(100)-wafers of the n-type (N(P)=2x10$^{14}$cm$^{-3}$) with a thin (curves 1), medium (curves 2) and thick (curves 3) oxide overlayer.

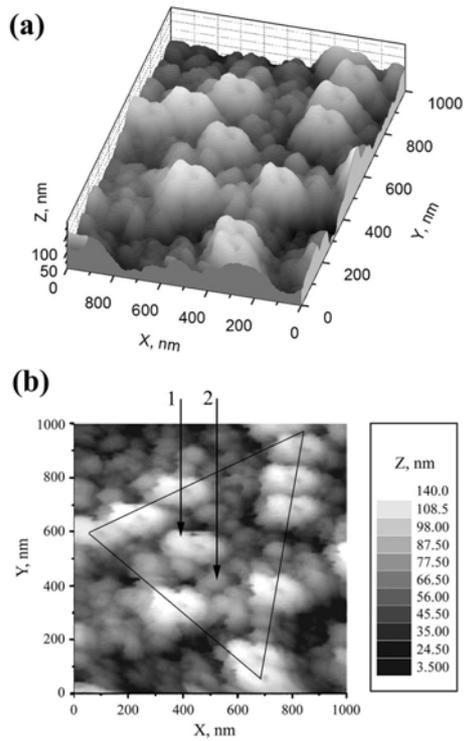

**Fig. 6.** STM image of the ultra-shallow boron diffusion profile prepared at diffusion temperature of 800°C into the Si(100) wafer covered previously by medium oxide overlayer. (a) - X∥[001], Y∥[010], Z∥[100]. Solid triangle and arrows that are labelled as 1 and 2 exhibit the microdefects with dimensions 740 nm, 225 nm and 68 nm, respectively, which are evidence of their fractal self-assembly (b).

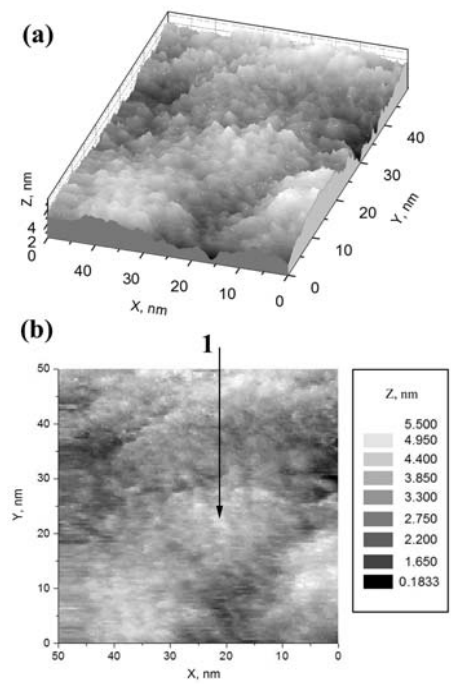

**Fig. 7.** STM image of the ultra-shallow boron diffusion profile prepared at diffusion temperature of 800°C into the Si(100) wafer covered previously by medium oxide overlayer. (a) - X∥[001], Y∥[010], Z∥[100]. The arrow labelled as 1 reveals the microdefect that is 20 nm across (b).

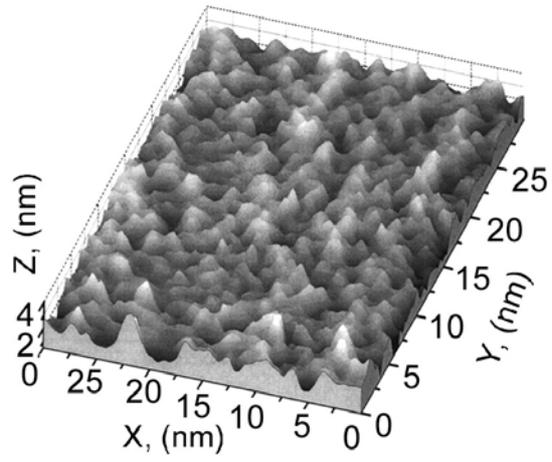

**Fig. 8.** STM image of the ultra-shallow boron diffusion profile prepared at diffusion temperature of 900°C into the Si(100) wafer covered previously by medium oxide overlayer. (a) - X||[001], Y||[010], Z||[100].

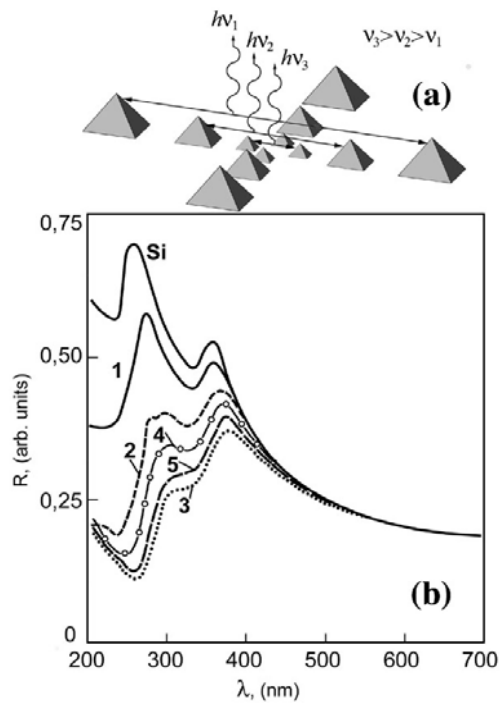

**Fig. 9.** (a) – The model of the self-assembled microcavity system formed by the microdefects of the fractal type on the Si(100) surface. (b) - The spectra of the reflectance from the n - type Si (100) surface and from the ultra-shallow boron diffusion profile prepared at the diffusion temperature on the n - type Si (100) surface covered previously by thin oxide overlayer: 1 - T=750°C, 2 - 800°C, 3 - 850°C, 4 - 900°C, 5 - 950°C.